\documentclass[]{spie}  %>>> use for US letter paper
\pdfoutput=1
\usepackage{amsmath,amsfonts,amssymb}
\usepackage{orcidlink}
\usepackage{comment}
\usepackage{xspace}
\usepackage{caption}
\usepackage{float}
\usepackage{longtable}
\usepackage{graphicx}
\usepackage{orcidlink}
\usepackage{placeins}
\usepackage{makecell}
\usepackage{array}
\usepackage{tabularx}
\usepackage{multirow}
\usepackage{listings}
\usepackage{lineno}
\usepackage{subfigure}
\usepackage{xstring}

\newcommand{\code}[1]{\texttt{#1}}
\newcommand{\ocs}[1]{\texttt{ocs}}

\newcommand{\hk}{\texttt{G3tHK}\xspace}

\newcommand{\daqsatp}{\texttt{daq-satp1}}
\newcommand{\crossbar}{\texttt{crossbar}}

\title{The Simons Observatory: Deployment and current configuration of the Observatory Control System for SAT-MF1 and data access software systems}

\author[1]{Sanah Bhimani\,\orcidlink{0000-0002-9763-1663}}
\author[1]{Jack Lashner\,\orcidlink{0000-0002-6522-6284}}
\author[2, 5]{Simone Aiola\, \orcidlink{0000-0002-1035-1854}}
\author[10]{Kevin T. Crowley\, \orcidlink{0000-0001-5068-1295}}
\author[3,4]{Nicholas Galitzki\,\orcidlink{0000-0001-7225-6679}}
\author[12]{Remington G. Gerras\, \orcidlink{0009-0009-0876-9168}}
\author[7,8]{Kathleen Harrington\,\orcidlink{0000-0003-1248-9563}}
\author[2]{Matthew Hasselfield\,\orcidlink{0000-0002-2408-9201}}
\author[9]{Alyssa Johnson\,\orcidlink{0000-0001-6567-5816}}
\author[1]{Brian J. Koopman\,\orcidlink{0000-0003-0744-2808}}
\author[11]{Hironobu Nakata\, \orcidlink{0000-0002-6300-1495}}
\author[1]{Laura Newburgh\,\orcidlink{0000-0002-7333-5552}}
\author[1]{David V. Nguyen\,\orcidlink{0000-0002-7575-8145}}
\author[10]{Michael J. Randall\, \orcidlink{0009-0009-9806-2317}}
\author[1]{Max Silva-Feaver\,\orcidlink{0000-0001-7480-4341}}

\affil[1]{Wright Laboratory, Department of Physics, Yale University, New Haven, CT 06511, USA}
\affil[2]{Center for Computational Astrophysics, Flatiron Institute, New York, NY 10010, USA}
\affil[3]{Department of Physics, University of Texas at Austin, Austin, TX, 78712, USA}
\affil[4]{Weinberg Institute for Theoretical Physics, Texas Center for Cosmology and Astroparticle Physics, Austin, TX 78712, USA}
\affil[5]{Joseph Henry Laboratories of Physics, Jadwin Hall, Princeton University, Princeton, NJ 08544, USA}
\affil[6]{Department of Physics, The University of Tokyo, Tokyo 113-0033, Japan}
\affil[7]{Argonne National Laboratory, High Energy Physics Division, Lemont, IL 60439, USA}
\affil[8]{University of Chicago, Department of Astronomy and Astrophysics, Chicago, IL 60637, USA}
\affil[9]{Department of Astronomy \& Astrophysics, University of California, San Diego, La Jolla, CA 92093, USA}
\affil[10]{Department of Physics, University of California, San Diego, La Jolla, CA 92093, USA}
\affil[11]{Department of Physics, Faculty of Science, Kyoto University, Kyoto 606-8502, Japan}
\affil[12]{University of Southern California, 825 Bloomwalk, Los Angeles, CA, 90089, USA}

\authorinfo{Further author information: (Send correspondence to Sanah Bhimani)\\ Sanah Bhimani: E-mail: sanah.bhimani@yale.edu}

\begin{document} 
\maketitle

\begin{abstract}
The Simons Observatory (SO) is a Cosmic Microwave Background experiment located in the Atacama Desert in Chile. SO consists of three small aperture telescopes (SATs) and one large aperture telescope (LAT) with a total of 60,000 detectors in six frequency bands.\cite{sogoals} As an observatory, SO encompasses hundreds of hardware components simultaneously running at different readout rates—all separate from its 60,000 detectors on-sky and their metadata. We provide an overview of commissioning SO’s data acquisition software system for SAT-MF1, the first SAT deployed to the Atacama site. Additionally, we share insights from deploying data access software for all four telescopes, detailing how performance limitations affected data loading and quality investigations, which led to site-compatible software improvements.

\end{abstract}


\section{INTRODUCTION}
\label{sec:intro}  % \label{} allows reference to this section

The Simons Observatory (SO) is a newly commissioned ground-based Cosmic Microwave Background (CMB) experiment in the Atacama Desert, Chile. The Simons Observatory features three 0.5\,m diameter Small Aperture Telescopes (SATs) and one 6\,m diameter Large Aperture Telescope (LAT), all designed for precision measurements of cosmological signals across a wide range of angular scales. Measurements from the LAT target a constraint on the sum of neutrino masses and the effective number of relativistic species, while the SAT aims to measure or constrain the primordial tensor-to-scalar ratio\cite{sogoals}. To accomplish these goals, the observatory fields approximately 60,000 cryogenic bolometers across the four telescopes, and requires a large set of diverse instruments for cryogenics, telescope platform operations, calibration, and instrument and observatory health. SO recently deployed all four telescopes to the site, starting with a SAT sensitive to two mid-frequency bands centered at 90 and 150 GHz, hereafter referred to as SAT-MF1. 

%Conducting observations with any observatory necessitates well-organized data acquisition, control, and monitoring for all instruments involved. For motorized telescopes such as SO’s, this includes following scientifically-motivated scan strategies, coordinated responses from various systems within the telescope to ensure seamless detector data collection, acquire ancillary instrument data, and perform ancillary maintenance activities. Although similar data acquisition and control software systems exist for other telescopes, they do not meet the specific requirements of SO. To address this, we have developed the Observatory Control System (OCS) to meet SO's data acquisition, control, and monitoring needs.

Similar to any observatory, SO requires a software system to acquire data and provide control and monitoring of all systems and instruments. This system must allow us to follow a designated scan-strategy, coordinate commanding and acquisition across multiple subsystems, and send alerts when subsystems are in an unhealthy state\cite{alarms}. We developed the Observatory Control System (\code{ocs}) to meet these data acquisition, control, and monitoring needs\cite{ocs}. In addition, accessing ancillary data, referred to as ``housekeeping'' (HK) data, is challenging due to SO's serialized, asynchronous data structure. To enable fast and efficient exploration of HK data for the detector analysis pipeline, we developed \hk, an SQLite database with an HK data loading mechanism.

In these proceedings, we focus on the software deployment for SAT-MF1. In Section \ref{sec:satocs}, we describe the \code{ocs} deployment for SAT-MF1, including the software and physical hardware systems for the platform and site. In Section \ref{sec:hk} we describe the deployment of \hk, its use cases, lessons learned, and future plans.

\section{SAT-MF1 \code{ocs} Software Deployment} \label{sec:satocs}
\begin{figure}[ht!]
    \centering
    \includegraphics[width=0.7\linewidth]{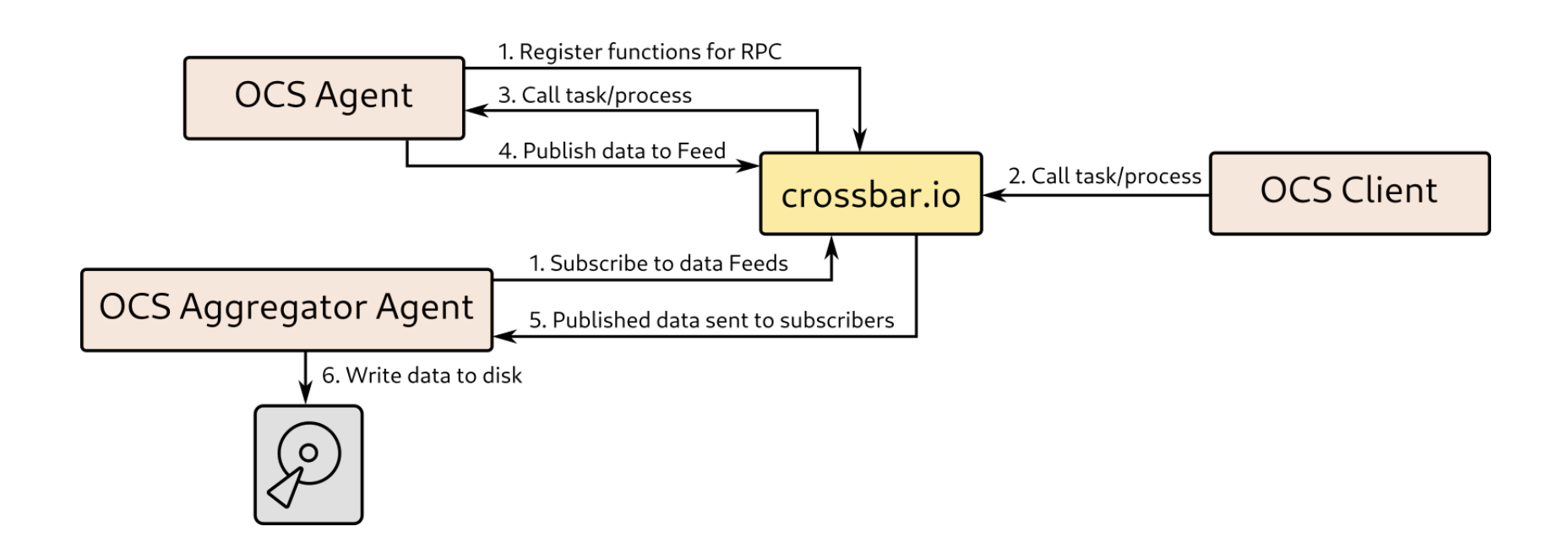}
    \caption{The \code{ocs} architecture, including major components, is shown. The \texttt{crossbar} router (yellow) facilitates communications between all other elements of the system. An \code{ocs} Client (right) is used to call a start, stop, or status for an operation. The operation is run by an \code{ocs} Agent (top left). Agents can facilitate \code{ocs} control and data acquisition for hardware units. If an operation collects data, that data is published to a Feed, which is accessed by an Aggregator Agent (bottom left). The Aggregator Agent writes the data to disk. Image by Brian Koopman\cite{ocs}.}
    \label{fig:ocsarch}
\end{figure}

\code{ocs} ensures commanding and data acquisition of the instrumentation, initiating and ending processes like telescope scanning, tuning detectors onto their superconducting transitions, or adjusting calibration equipment. The \code{ocs} architecture comprises two primary components: agents and clients, as shown in Figure~\ref{fig:ocsarch}. \code{ocs} agents are long-running software servers that interface with hardware or software components to coordinate data acquisition\cite{ocs}. Agents have operations classified as Tasks and Processes:
\begin{itemize}
\item \textbf{Tasks} run for a finite amount of time and have a natural completion, such as moving the telescope platform or setting the heater range for servoing a temperature sensor in a cryostat.
\item \textbf{Processes}, like the continuous “acq” process for data acquisition, run until stopped. \cite{ocs} 
\end{itemize}
\texttt{ocs} clients are scripts that orchestrate the actions of one or more \texttt{ocs} agents over the network. For example, an \texttt{ocs} client can instruct a managing agent to move a telescope platform to a specific position, after which the agent executes the command and returns the final position data. Crossbar.io facilitates  interactions between agents and clients. In addition, this system provides a real-time visualization for slow rate data via Grafana\cite{ocs}, as described in Section \ref{sec:satagents}. Detailed specifics on the alerts based on this system are described in Nguyen et al. 2024.\cite{alarms}

In this section, we describe the deployment of \code{ocs} for the SAT-MF1 platform and receiver, the first receiver deployed in SO. We begin with a discussion of the agents required for the SAT-MF1 receiver, platform, and site. We also include where the agents are running, and provide a discussion of the user interfaces and real-time visualization for the system. 

% commented out the longer details 
\subsection{SAT-MF1 Agents}\label{sec:satagents}

SAT-MF1 is one of two mid-frequency SATs and the first to begin observations, which commenced in October 2023. It consists of approximately 12,000 transition-edge sensor (TES) bolometers cooled to 90mK,\cite{galitzki2024simons, Zhu_2021, Gudmundsson_2021, McCarrick_2021} a spinning cryogenic half-wave plate at 1K temperatures, and an extensive set of housekeeping components.

\FloatBarrier Currently, there are a total of 69 SAT-MF1 \code{ocs} agents running on-site. Of these, 8 are `core \code{ocs}' agents: the Aggregator, the Registry, the InfluxDB Publisher, and 5 Host Manager agents. Detailed descriptions of these agents can be found in Koopman et. al. 2024, but we briefly describe each for context:

%and the sequencer agents. 

\begin{itemize}
\item Aggregator -- subscribes to `slow' data rate feeds, referred to here as housekeeping (HK) data, and writes to disk. Slow data comes from all components that are not the fast-cadence (200\,Hz) detector data.  
\item Registry -- monitors all active agents, tracking the status of each agent's Tasks and Processes.
\item InfluxDB Publisher -- similar to the Aggregator agent, but publishes the HK data to an InfluxDB server for viewing on Grafana\cite{ocs} instead of writing to disk.
\item The Host Manager -- provides a way to start and stop agents running on a given machine or host.
\end{itemize}

%which are more clearly defined in Section \ref{sec:nodes} 
%\textbf{host is not defined}

The remaining 61 agents within the SAT-MF1 \code{ocs} system are integral to the operation of the SAT's receiver and telescope platform: the telescope control unit (ACU); the cryogenics (Bluefors dilution refrigerator, thermometry, vacuum components); cryogenic half-wave-plate and wiregrid calibrators; and SMuRF (detector readout) agents for monitoring, controlling, and data transfer. 

Additionally, the site \code{ocs} system and supporting software infrastructure facilitate remote operations for the entire observatory. Separate from SAT-MF1's \code{ocs} setup, there are 7 additional site agents crucial for assessing observatory conditions and ensuring effective telescope operations, noted in Table \ref{tab:sitetab}: 5 generator agents that monitor fuel levels for the site's diesel generators; a radiometer agent that monitors atmospheric precipitable water vapor (PWV) to assess whether conditions are suitable for observations on any given day; and a weather monitor that tracks local site conditions such as wind speed, wind direction, humidity, and more. %The Bluefors Agent queries the logs generated by the Bluefors software, and records a variety of data (helium pressures and flows, temperatures, vaccuum pressure, valve states)

The salient features of SAT-MF1 agents and their data characteristics are summarized in Table \ref{tab:ocstab}, including data rate, number of fields, data type, and the development origin of each agent. In total, the 76 agents (69 SAT-MF1 + 7 site) generate data at a rate of approximately 3.74 Mbps. This data rate includes 2074 \code{ocs} fields, each representing a specific data component collected by an agent, such as temperature, voltage, or status information. For example, both temperature and voltage readings for each of the 11 Lakeshore 372 channels would be represented by 22 separate \code{ocs} fields.
Of the 2074 fields, SAT-MF1 contributes 1524 fields with a data rate of 3.58 Mbps, while components of the site \code{ocs} sub-system mentioned in this proceedings contribute 550 fields with a data rate of 0.162 Mbps. Many of the agents described in Table \ref{tab:ocstab} were developed by hardware experts across the SO collaboration, aligning with one of the primary goals of the \code{ocs} architecture.

\begin{figure}[ht]
    \centering
    %\hspace*{-1.5cm}
    \includegraphics[width = \linewidth]{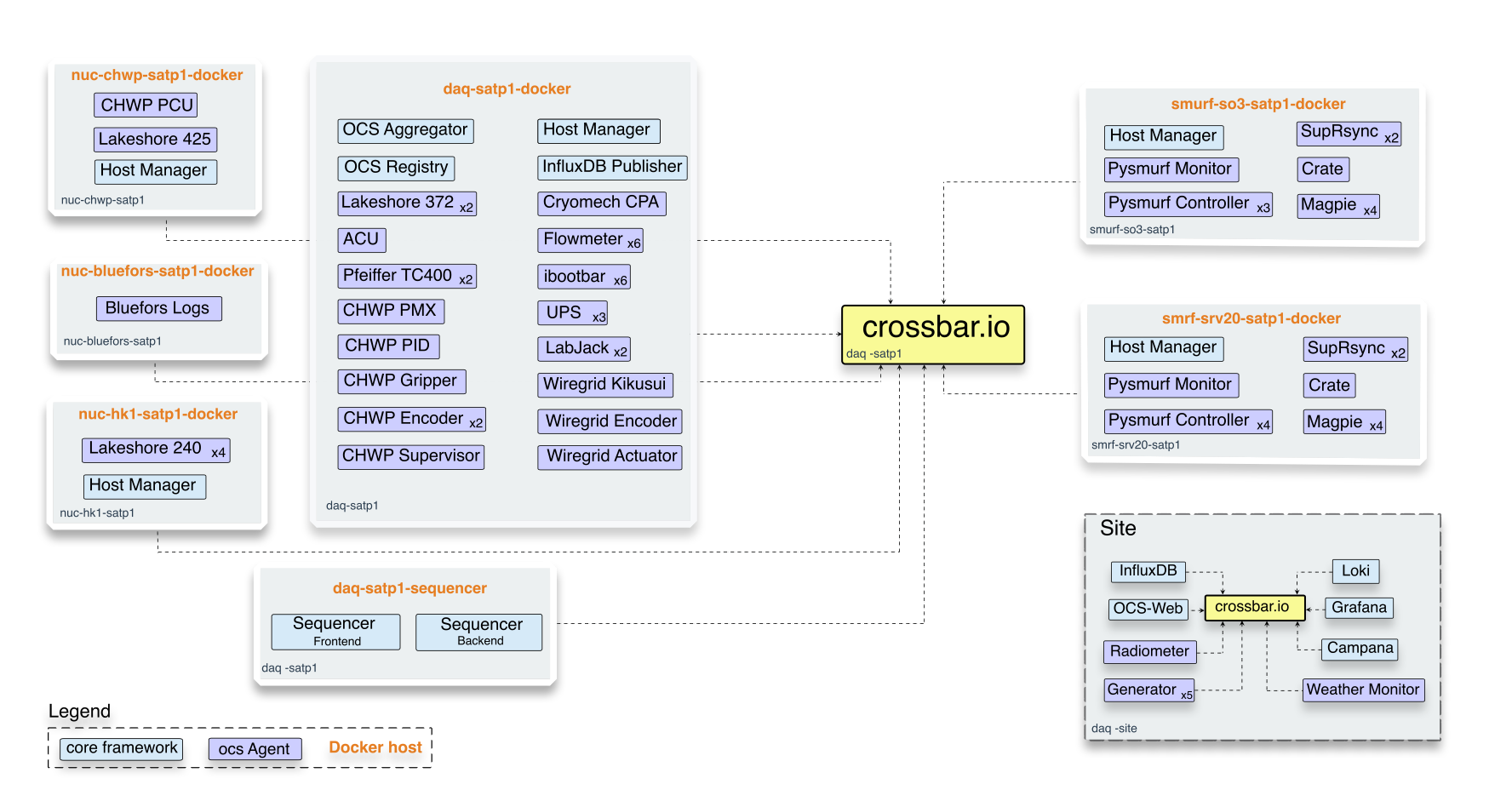}
    \caption{\code{ocs} deployment for SAT-MF1 at the site, referred to in the software as \code{satp1}. This layout breaks down the distribution of agents across the instrument by host, further detailed in Section \ref{sec:nodes}. A subset of the site software system, including \code{ocs}, ocs-web, and open-source services is shown on the bottom right. Site agents monitor site-wide hardware relevant to observatory operations, which includes monitoring precipitable water vapor (PWV) values, weather conditions, and generator fuel levels. Details on the open-source software and ocs-web can be found in Koopman et al. 2020.}
    \label{fig:satp1ocs}
\end{figure}

Figure \ref{fig:satp1ocs} provides the current layout for SAT-MF1 HK agents, including a subset of the site software system on the bottom right. The site setup provides the framework for the observatory for alerting (Campana\cite{alarms}), commanding (ocs-web), and monitoring \code{ocs} data and logs (Grafana, Loki)\cite{ocs24}. For details surrounding the complete site layout, refer to Koopman et al. 2024.

The fast-cadence data from the detectors is acquired separately by the SMuRF readout system \cite{smurf, Kernasovskiy_2018}. The SMuRF electronics readout is managed by four \code{ocs} agents: Pysmurf Monitor, Pysmurf Controller, Crate, and Magpie. Since real-time monitoring of detector streams is valuable but unfeasible with our `slow' HK acquisition streams, the Pysmurf Monitor Agent can send a small subset of downsampled detector timestreams to the HK aggregation system for viewing through Grafana.

\begin{table}[H]
    \centering
    \begin{tabular}{|l|p{3cm}|p{2.2cm}|p{1.5cm}|p{2.3cm}|p{2.2cm}|}
    \hline
    \textbf{Subsystem} & \textbf{Site \code{ocs} Agent} & \textbf{Data Rate (kbps)} & \textbf{No. of Fields} & \textbf{Data Type} & \textbf{Where \newline Developed} \\
    \hline
    \multirow{3}{*}{Environment} & UCSC Radiometer & 0.015 & 1 & Float & DAQ \\
    & Generator & 160.4  & 474 & Float, Int & DAQ \\
    & Weather Monitor & 1.8  & 75 & Float, Int & DAQ \\
    \hline
    \textbf{Total} & -- &\textbf{0.162 Mbps} & \textbf{550} & -- & -- \\ %0.4675 megabytes/sec y
    \hline
    \end{tabular}
    \vspace{2mm}
    \captionsetup{width=\textwidth}
    \caption{A table describing data characteristics of a small set of \code{ocs} agents for the site.  The total data rate for this system is 0.162 Mbps, with 550 fields across 7 agents.}
    \label{tab:sitetab}
\end{table}

\begin{table}[H]
    \centering
    \begin{tabular}{|l|p{3cm}|p{2.2cm}|p{1.5cm}|p{2.3cm}|p{2.2cm}|}
    \hline
    \textbf{Subsystem} & \textbf{SAT-MF1 \newline \code{ocs} Agent} & \textbf{Data Rate (kbps)} & \textbf{No. of Fields} & \textbf{Data Type} & \textbf{Where \newline Developed} \\
    \hline
    Telescope & ACU & 2520  & 197 & Float & DAQ \\
    \hline
    \multirow{8}{*}{Cryo} & Bluefors & 288 & 45 & Float & DAQ \\
    & Pfeiffer TC400 & 0.384  & 6 & Float & UCSD \\
    & Cryomech CPA & 5.44 & 34 & Float & UCSD \\
    & Lakeshore 372 & 7.04 & 11 & Float & DAQ \\
    & Lakeshore 240 & 8.32 & 52 & Float & DAQ \\
    & Lakeshore 425 & 0.064 & 1 & Float & UTokyo \\
    & Flowmeter & 0.015 & 12 & Float & DAQ \\
    \hline
    Labjack & Labjack & 7.42 & 80 & Float & UCSD \\
    \hline
    \multirow{2}{*}{Power} & ibootbar & 1.24 & 144 & Float, Int, Str & DAQ \\
    & UPS & 0.91 & 105 & Float, Int, Str & DAQ \\
    \hline
    \multirow{8}{*}{Calibration} & HWP PCU & 0.024 & 1 & Str & UTokyo \\
    & HWP PID & 1.28 & 4 & Int & UCSD/UTokyo \\
    & HWP PMX & 5.312 & 6 & Float, Int, Str & UTokyo \\
    & HWP Gripper & 0.31 & 24 & Float, Int & DAQ/UCSD \\
    & HWP Encoder & 10.8 & 42 & Float, Int & UTokyo \\
    & Wiregrid Encoder & 4.16 & 13 & Float, Int & Kyoto Uni. \\
    & Wiregrid Kikusui & 0.32 & 5 & Float & Kyoto Uni. \\
    & Wiregrid Actuator & 0.83 & 13 & Int & Kyoto Uni. \\
    \hline
    \multirow{4}{*}{\code{ocs} Core} & Registry & 157.3 & 365 & Float, Int, Str & DAQ \\
    & Aggregator & -- & -- & -- & DAQ \\
    & Host Manager & -- & -- & -- & DAQ \\
    & InfluxDB Publisher & -- & -- & -- & DAQ \\
    \hline
    \multirow{4}{*}{Smurf} & Pysmurf Monitor & 8.512 & 133 & Float & Detector/DAQ \\
    & SupRsync & 1.728  & 20 & Float, Int, Str & Detector/DAQ \\
    & Pysmurf Controller & 544.32 & 126 & Float, Int, Str & Detector/DAQ \\
    & Crate & 2.016 & 84 & Str & Detector/DAQ \\
    \hline
    \textbf{Total} & -- &\textbf{3.58 Mbps} & \textbf{1524} & -- & -- \\ %0.4675 megabytes/sec y
    \hline
    \end{tabular}
    \vspace{2mm}
    \captionsetup{width=\textwidth}
    \caption{A table describing data characteristics of \code{ocs} agents setup for SAT-MF1, divided by subsystem. The number of fields represents those currently online for all instances of that agent for SAT-MF1. For instance, the 12 flowmeter data fields in the cryo subsystem correspond to 6 active flowmeter agents, each with 2 fields (flow in liters/min and temperature in Celsius). The total data rate for the instrument is 3.58 Mbps, with 1524 query-able data fields from the telescope's 69 agents. The registry agent, described in Section \ref{sec:satagents}, monitors all active agents and the status of their corresponding Tasks and Processes. It currently tracks 365 fields, contributing the most to the data rate. Data rates for other core \code{ocs} agents are not noted, as their characteristics do not require data fields being written to disk.}
    \label{tab:ocstab}

\end{table}

 % \footnote{smurf numbers are wrong: there exists 47non smurfrelated agents, 7 psymurf controller agents, 2 pysmurf monitor agents 4 suprsync agents+ 7 magpie agents + 2 crate agents} 

\subsection{User Interfaces}

\begin{figure}[ht]
    \centering
    \includegraphics[width =0.9\linewidth]{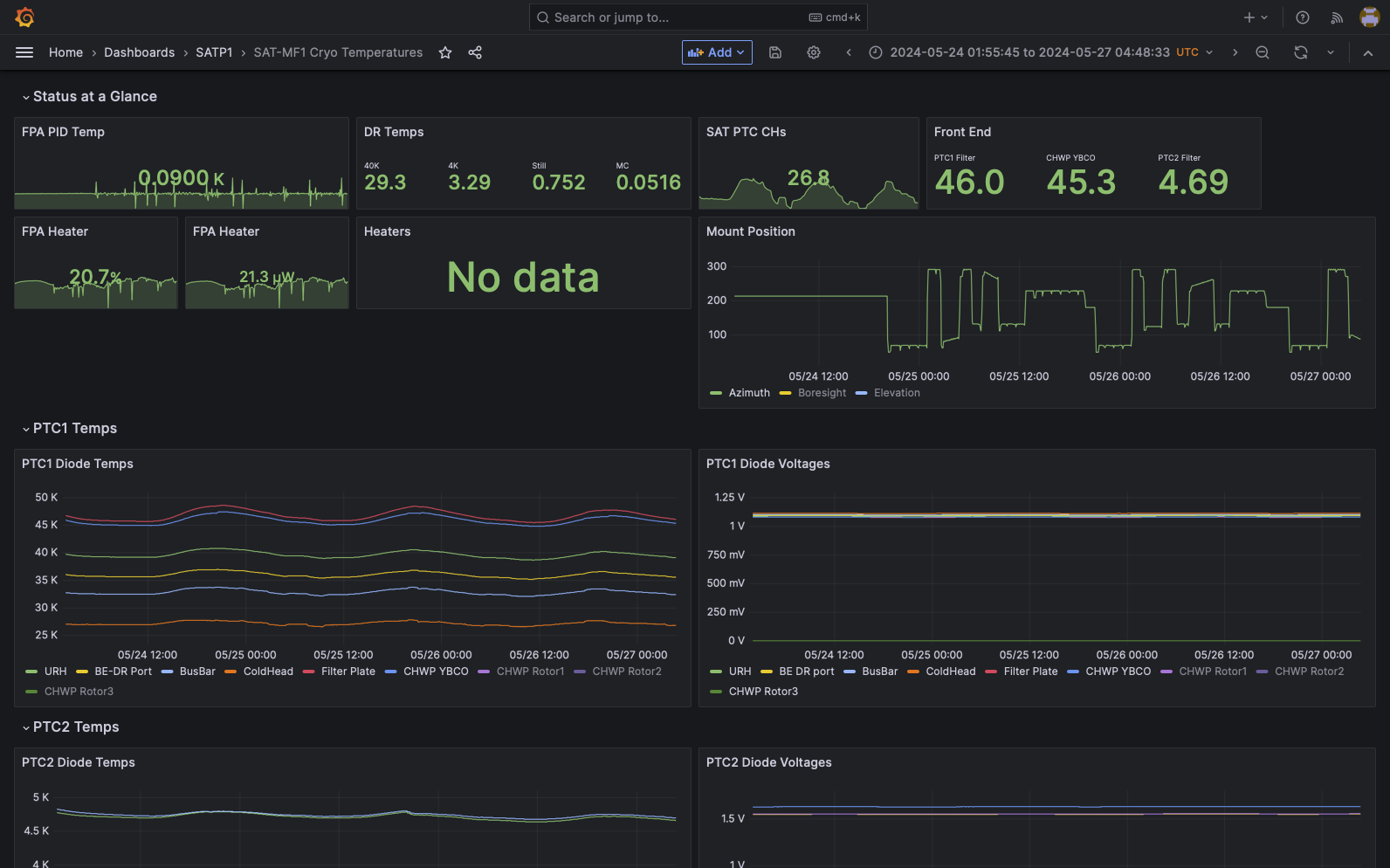}
    \vspace{0.5mm}
    \caption{Screenshot of the the SAT-MF1 Grafana dashboard for monitoring the state of the dilution refrigerator and cryogenic temperatures, as well as brief information about the state of the telescope platform.}
    \label{fig:grafana}
\end{figure}

\begin{figure}[ht]
    \centering
    \includegraphics[width = 0.9\linewidth]{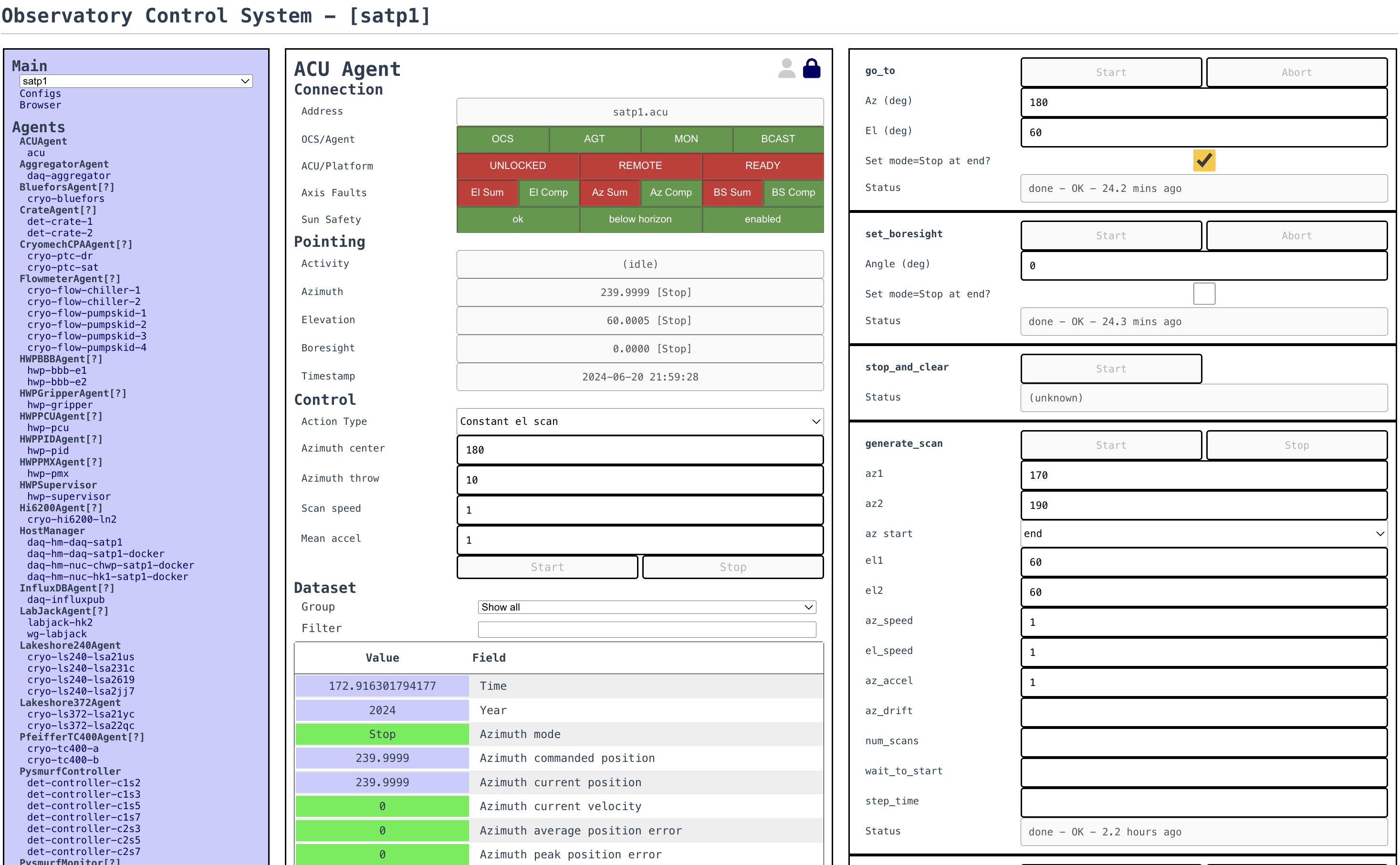}%{satmf1ocswebacu.png}
    \vspace{0.5mm}
    \caption{Screenshot of ocs-web for SAT-MF1's telescope platform agent, the ACU. The right-hand side displays several tasks a user can command the ACU to perform via ocs-web. For example, if weather conditions deteriorate quickly during a scheduled observation, a remote observer will use the \lstinline{go_to}
    Task to move the ACU to weather-safe azimuth and elevation positions. Entering parameters for \lstinline{go_to} on the ocs-web panel runs an \code{ocs} Client.}
    \label{fig:ocsweb}
\end{figure}

The tools and interfaces used by and designed for SO provide essential functions for monitoring, commanding, and visualizing the status and data of various HK agents. Below is an overview of the user interface with \code{ocs} and supporting software for SAT-MF1, now fully operational and observing:

\begin{itemize}
\item The `sequencer' -- a tool with a graphical front end designed to run \code{ocs} clients\cite{ocs24}. It utilizes another library to enable the sequential command of multiple clients, ensuring efficient and coordinated operations.\cite{ocs24, nextline}
\item Grafana -- open-source time series display for live monitoring, see Figure \ref{fig:grafana}.
\item ocs-web -- a collection of \code{ocs} clients, presented as a graphical user interface in a web browser and implemented using JavaScript, is primarily used for direct interaction with a given agent for specific commands\cite{ocs24}; see Figure \ref{fig:ocsweb}.

\item Campana -- an alarms and alerts system\cite{alarms} based on Grafana.

\item Loki -- a log aggregation tool developed by Grafana Labs, allowing users to follow real-time logs or query past logs directly within the Grafana interface.
\end{itemize}

A more comprehensive description on the supporting software infrastructure can be found in Koopman et al. 2024; more details on the alarms and alerting system can be found in Nguyen et al. 2024.

Since commissioning began, both on-site and remote users have constantly interacted with \code{ocs} at the site. SO uses Remote Observing Coordinators (ROCs) to remotely monitor the telescope and follow the health of the instrument. Now transitioning to the operations phase, ROCs follow a routine set of actions, typically performing the following tasks during each shift:

\begin{enumerate}
    \item \textbf{Check PWV Values:} First, the ROC checks water vapor conditions on Grafana for a real-time weather estimate. If the values are favorable (less than 2mm), they proceed to the next step.

    \item \textbf{Deploy Schedule:} The ROC deploys a pre-made schedule (sequence of commands) onto \texttt{nextline}\cite{nextline}, the front-end feature of the sequencer. \code{nextline} runs the schedule.

    \item \textbf{Monitor Alarms:} While the schedule runs, the ROC monitors the Grafana alarms page. If any alarms go off, they receive notifications via text, email, and/or phone (using Campana to choose their preferred notification methods\cite{alarms}). They then check the status of the specific alarm on Grafana. 

    \item \textbf{Check Agent Logs if Necessary:} For agent communication-related alarms, the ROC may use Loki on Grafana to view logs of the failing agent.

    \item \textbf{Pause Schedule:} If the alarm is severe enough, the ROC needs to put the telescope in a specific state, which requires pausing the deployed schedule. This is done using the \code{nextline} front-end feature. 

    \item \textbf{Command Hardware:} After the schedule is paused, the ROC commands the telescope using ocs-web. This can include commanding instruments like the cryogenic half-wave plate, thermometry readout devices, and/or the telescope ACU. By clicking any \code{ocs}  Task on the right-hand side of an agent panel in ocs-web, they effectively run an \code{ocs} client. See Figure \ref{fig:ocsweb}.

    \item \textbf{Handle Site Communications:} Even with telescope operations running smoothly, the ROC may receive notifications from SO's remote operations channel (\texttt{satp1-remote-ops} on Slack for SAT-MF1). This may require pausing the schedule and changing the state of the instrument for site testing.
    
\end{enumerate}

\subsection{Nodes and Hosts Setup}\label{sec:nodes}

The \code{ocs} agents run on computing resources at the site. Many agents can communicate via HTTP or TCP interfaces, allowing them to run on any computer within the site network. Each telescope platform is assigned a virtual machine hosted on a Linux server\cite{ocs24}. In this proceedings, we will refer to this virtual machine, or any similar computing unit within a network, as a node.

%---whether it is a physical or virtual machine.

%Each telescope platform is assigned a virtual machine hosted on a Linux server\cite{ocs24}.  In this proceedings, we will refer to such a machine or a virtual machine as a node---a computing unit within a network, which can either be a physical or virtual machine.

%Each telescope platform has a designated virtual machine on a Linux server\cite{ocs24}, which is located in the site office.
\texttt{daq-satp1} serves as the dedicated site-managed node for SAT-MF1, running 36 out of the 69 agents. Interface limitations or impracticality of network communication for some hardware components required nodes to be physically installed closer to the hardware. Since \code{daq-satp1} is hosted on a Linux server that is located in the site office, it could not host agents for these types of hardware, necessitating the integration of additional nodes into SAT-MF1's \code{ocs} system.
Most other nodes on the platform are Intel NUCs\footnote{https://simplynuc.com/onyx/} (mini-PCs), although can also be anything with a processor (BeagleBone\footnote{https://www.beagleboard.org}, Raspberry Pi\footnote{https://www.raspberrypi.com}, etc.). The \crossbar \ router is configured on \texttt{daq-satp1}, allowing all agents, whether on \texttt{daq-satp1} or other nodes, to connect to it\cite{ocs24}. A dedicated port on the router ensures that SAT-MF1 operations remain isolated from those of other telescopes.

%\textbf{SB Note:} Still need to add how many agents of 69 come from the SMuRF servers

\subsubsection{SAT-MF1 Hosts}
The agents described in Section \ref{sec:satagents} can be grouped by hosts, denoted in Figure \ref{fig:satp1ocs}. A host is a physical or virtual environment that provides the necessary infrastructure for running various applications and services. It can be a bare-metal server or a virtualized instance, such as a Docker host, offering computational resources, storage, and network connectivity. A host may manage multiple containers, each functioning as an independent computing environment within the larger system. We predominantly use Docker for deploying \code{ocs} within SO as it allows us to package the software, libraries, and configuration files for each agent into distributable containers, creating isolated and reproducible environments that are invaluable for site deployment.

With respect to SAT-MF1's \code{ocs} infrastructure (see Figure \ref{fig:satp1ocs}):
\begin{itemize}
    \item \code{daq-satp1} is a node,
    \item \code{daq-satp1-docker} is the Docker host on \code{daq-satp1}. This host is where the 36 agents on \daqsatp\ reside.
    \item  each agent within the \code{dat-satp1-docker} host is its own container, i.e., 36 containers
\end{itemize}

This structure applies across all nodes and hosts within the SAT-MF1 system.
 
 In Figure \ref{fig:satp1ocs}, we see that \code{daq-satp1} also contains a second Docker host that manages the sequencer containers, separate from the 36 agents. In addition, there is a third Docker host (not pictured) on \daqsatp, which manages the \crossbar \ Docker container to house the router that interfaces with all of SAT-MF1's agents. It is isolated from any host on the network that contains the agents. This separation ensures that restarting SAT-MF1-specific hardware and agents will not disrupt long-term services like crossbar.io.\cite{ocs}

\subsubsection{Enclosures and Node Setup}
SAT-MF1 houses detector readout and housekeeping hardware in an enclosure system of 12 units. These enclosures are mounted on the telescope and are categorized as: SMuRF readout enclosures, cold half-wave plate and wiregrid enclosures, the azimuth enclosure, the coolant control enclosure, thermometer readout enclosures, the gas handling system enclosure, and the turbo pump enclosure. They also include power supplies such as iBootBars and UPS units, as well as timing hardware. Table \ref{tab:enclosures} includes a list of SAT-MF1 nodes at the site and their corresponding enclosures. 

%Some hardware components require a USB or direct connection to a node to connect to the network. This necessitated mounting a computing unit (NUC or similar) within the enclosures to integrate them into the \code{ocs} system.

The NUC computer housing the \code{nuc-chwp-satp1-docker} host for the cold half-wave plate (CHWP) readout and the NUC containing the \code{nuc-hk1-satp1-docker} host for thermometry readout via the Lakeshore Cryogenics devices (Lakeshore 240s) are both on Linux machines and operate similarly to the larger \code{daq-satp1-docker} host.

%\footnote{\textbf{SB note to address:} is it too much to mention that this is because the require USB connections and hence, nucs on the platform}

The Bluefors NUC is a Windows computer running the Bluefors 3He/4He LD400 dilution refrigerator readout. The Bluefors agent queries the logs generated by the Bluefors software on the NUC and publishes the data to both the aggregator agent and InfluxDB for Grafana visualization and alarms. Note that this Bluefors setup runs within a Docker container, but is the only one without a Host Manager agent, as the Host Manager is not Windows-compatible.

All three NUCs---regardless of OS---publish feeds, query data, command agents as applicable, and write data to disk via the aggregator agent through the \crossbar \ connection on \daqsatp.

\subsection{Enhanced Site Deployment Structures}

One goal while developing \code{ocs} was for the agents to be developed by hardware experts in the lab, such that deploying the software required minimal changes from their lab experience. Nonetheless, deployment to the site did include a few notable changes: the use of a centralized configuration file management system\cite{ocs24}; the integration of the Host Manager agent\cite{ocs, ocs24}; expansion of agents covered by ocs-web, and now the primary use for commanding each individual agent at the site; significant refactoring of critical calibration agents; and the implementation of Campana for alarms and alerts\cite{alarms}.

Site configuration files (SCFs) are a core component to the deployment of \texttt{ocs} agents---configuration files are required on each computer running \texttt{ocs}. A SCF contains information such as the \crossbar \ router address, a list of agents grouped by host, and a unique `instance-id' for each agent instance to differentiate between multiple copies (e.g., four Lakeshore 240s). Prior to site deployment, SCFs were managed by each lab developing SO's receivers. At the site, SCF management requires telescope experts to push to a private GitHub repository, \texttt{ocs-deployment-configs} which uses a GitHub Actions runner that executes an Ansible playbook to deploy SCF changes to the site computers. This allows the configuration file to be automatically deployed and updated when a new agent has been added or when an existing agent parameter is added to the SCF.

%an update to an existing parameter is desired. 
While critical agents for the telescopes were initially developed and integrated in SO labs before commissioning, new agent development or updates to existing agents are not restricted to off-site locations. A \code{daq-dev} node exists to enable agent development at the site and to test directly on the site hardware\cite{ocs24}. This node allows for software updates to be tested without introducing test data sets into the full SO data stream. 

Agents for the half-wave plate were refactored and tested during lab integration to find an optimal setup for the site, resulting in the breakdown of calibrator actions into separate, related half-wave plate agents. 

Campana has proven indispensable for determining the exact issues when alarms sound, which is crucial in a large observatory. For further information, refer to Nguyen et al., 2024.

\section{Housekeeping Data Access Tools}\label{sec:hk}

At the time of writing, across the entire observatory (four telescopes + the site \code{ocs}  systems), we have over 250 agents and over 5000 slow-rate housekeeping data fields. It is effective to have a system that allows efficient investigation of these fields without the need to repeatedly read full data files for each analysis. In this section, we briefly describe the data formats used for SO, a database we developed to query HK data, and highlight some of the most important applications of this database. 

\subsection{Data packaging}

To capture all observatory data, the Simons Observatory has adopted the .g3 file format, originally developed by the South Pole Telescope as part of their \texttt{spt3g} software\footnote{\url{https://cmb-s4.github.io/spt3g_software/}}. 
A .g3 file is a series of \textit{frames}, where each frame is a mapping from strings to custom serializable data types. Each frame typically contains either metadata, or data taken by a data provider over a period of time. The format has been adapted to meet SO's specific requirements and is referred to as \texttt{so3g}. For SO, the data files are heterogeneous: housekeeping files are written hourly, while detector .g3 files are written every 10 minutes, with specific detector setup operations recorded separately from the detector timestream data. For SO, typical CMB observations are $\sim$65 minutes long whereas planet scans can last approximately 1.5 to 3 hours,  further highlighting the variability in data file durations. Hence, SO's data packaging follows three levels of data production:

\begin{itemize}
    \item Level 1: Initial acquisition of data into memory.
    \item Level 2: Level 1 data are written to disk, with a specific folder structure, becoming `aggregated data' known as .g3 files.
    \item Level 3: Data files are grouped together into the Level 3 format, hereafter known as books. Books are essentially directories of .g3 files which have been re-framed and reorganized to reflect detector observations for user-friendly analysis. Books are our final storage format, to be used in analysis both at the site and off-site. For more details on data packaging, refer to Guan et al., 2024 ~\cite{yilun_spie}.
\end{itemize}

While the reshuffling from .g3 files to books is critical for organizing detector data by telescope observation (e.g., Planet scan vs. CMB scan), housekeeping books remain unchanged from their Level 2 .g3 counterparts. 

\subsection{\hk Database}
% (data loading will be described later)

We developed \hk, which consists of two components: a database and a data loader. \hk uses an SQLite database and organizes data into three tables: .g3 files, HK \code{ocs} agents, and the agents’ corresponding fields. For example, for SAT-MF1, a given .g3 HK file has data from all 69 agents, and 1524 fields, as described in Section~\ref{sec:satagents}. The .g3 files table stores brief information about each HK file with attributes such as: unique file id, filename, file start time, file location, and source of information (i.e., satp1, satp2, satp3, lat, site); see Figure~\ref{fig:hkfiles}. 

\begin{figure}[ht]
    \centering
    \includegraphics[width = 0.9\linewidth]{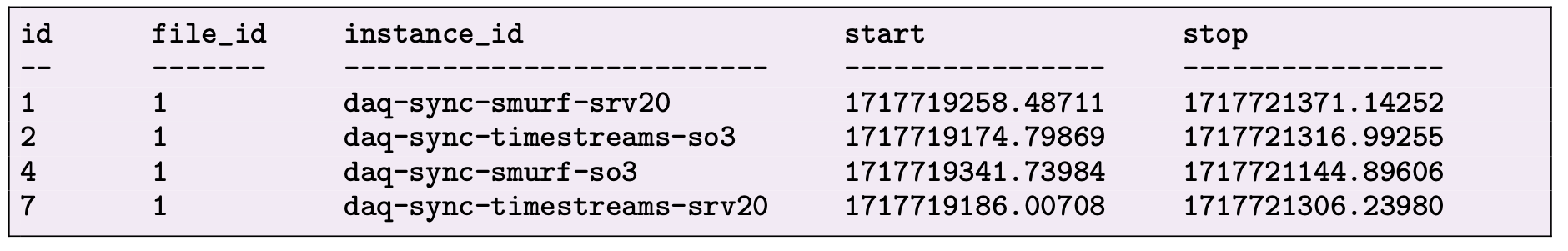}
    \vspace{0.5mm}
    \captionsetup{width=0.9\linewidth}
    \caption{An example of an \hk \ \  HK agents table. The instance-id's are unique identifiers for each agent instance for the suprsync agent, where there is a separate agent instance for detector timestreams and detector metadata.}
    \label{fig:hkagents}
\end{figure}

The HK agents table in Figure \ref{fig:hkagents} contains information about each agent within an HK file. Attributes in this table include: the same file ID in the .g3 files table, a unique agent ID number, the instance-id, and file start and end timestamps. The HK fields table lists all fields for each agent in every HK file with attributes including the unique file ID, the unique agent ID, field name, and field start and end time (which may be different from the file start and end time). We note that this database is populated for every file, regardless of observation or detector status, thus keeping a record of all HK data recorded at the site, even if we were not observing the sky.

There are two separate instances of this database structure. One is used by the bookbinder, as described in Section \ref{sec:bookbinder}, to transform Level 2 data to Level 3. This database contains only the information relevant for data packaging from Level 2 to Level 3. For SAT-MF1, this means the database includes 153 fields per HK .g3 file (compared to 1524), reducing the need to query a much larger database during the bookbinding process. The second database instance is used for HK data exploration and is streamlined to include only the agents most commonly queried for contextualizing detector behavior. It primarily excludes information from core \code{ocs} agents and power subsystem agents, such as the ibootbar and UPS, which display extensive status information on power supplies.

%The second database instance is used for HK data exploration. This is also reduced, to include only agents that are typically/most commonly queried for data investigation. Mainly the agent information removed are OCS core agent information or site-related status indicators (ibootbar status, etc). 

\subsubsection{For Bookbinding}\label{sec:bookbinder} %or 
% specify that this database only contains information about agents that are relevant for datapackagin
% this is critical for bookbinding; read Yilun's for bookbinding info for help
SAT-MF1 contains approximately 12,000 detectors distributed across 7 silicon wafers, with .g3 data files organized by wafer. The bookbinder requires precise start and end times for each wafer, as well as information on which wafers were streaming data, to re-frame the detector data. These times are obtained from the .g3 file itself and stored in the \hk instance developed for the bookbinder.

\begin{figure}[ht]
    \centering
    \includegraphics[width = \linewidth]{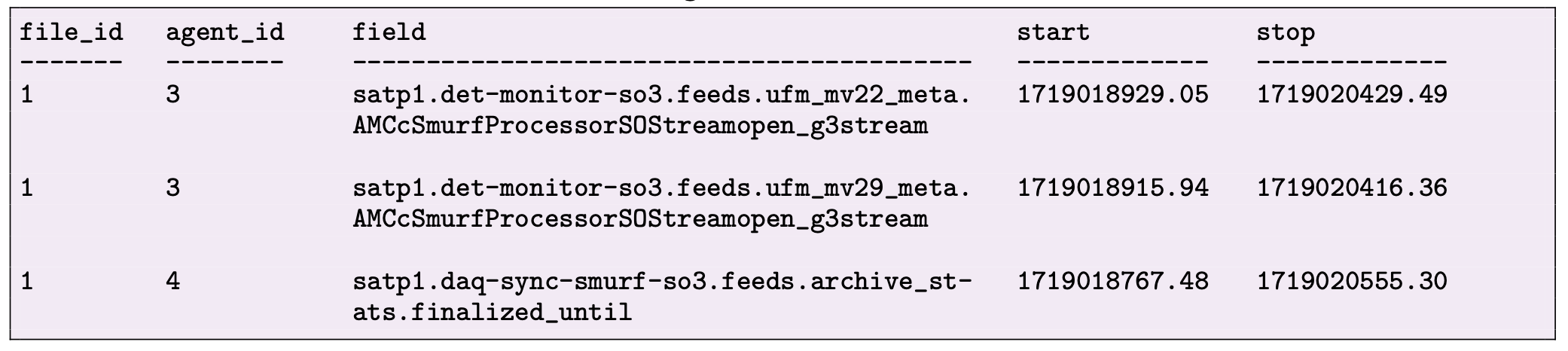}
    \caption{An example of a \hk HK fields table is shown, listing two fields from the \code{daq-monitor-so3} pysmurf monitor and one \code{daq-sync-smurf-so3} suprsync agent. The field names containing \lstinline{AMCcSmurfProcessorSOStreamopen_g3stream} have start and stop times that indicate which detector wafers were active and streaming data. The field \lstinline{daq-sync-smurfso3.feeds.archive_stats.finalized_until} indicates the last time the suprsync agent for a SMuRF server successfully checked the files and copied them to \code{daq-satp1}, making them available for bookbinding. }
    \label{fig:hkfields}
\end{figure}

\begin{figure}[ht]
    \centering
    \includegraphics[width = \linewidth]{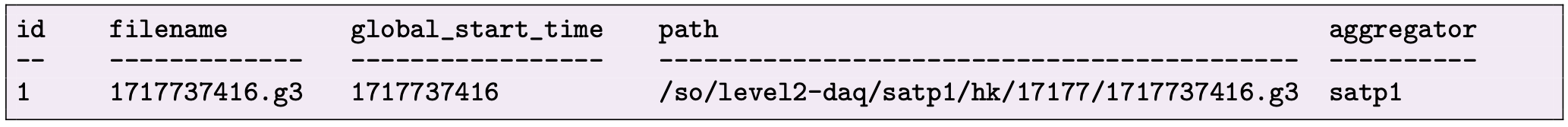}%{satmf1ocswebacu.png}
    \vspace{0.5mm}
    \caption{A \hk HK files table. The \lstinline{global_start_time} is derived from the ctime in the filename and is provided as a separate column in the SQLite table to facilitate queries for files within a specific time range. In this table, SAT-MF1 is noted as \code{satp1} under the \code{aggregator} column. At the observatory site, the aggregator serves as the identifier for the agents assigned to each telescope, as well as the site system. The aggregator name is also part of the full data field, as shown in Figure \ref{fig:hkfields}. }
    \label{fig:hkfiles}
\end{figure}

In Figure \ref{fig:hkfields}, two of the seven wafers for SAT-MF1 are shown: mv22 and mv29, along with their corresponding streaming times in a single hour long HK file. The bookbinder uses the \hk instance by querying the pysmurf monitor for any of the streaming wafers to accurately reframe the .g3 files. The pysmurf monitor tracks when each wafer was streaming by using the field that contains \lstinline{AMCcSmurfProcessorSOStreamopen_g3stream}, which indicates which pysmurf monitors were in control of which wafers during specific periods. This ensures that timestreams and metadata from each wafer are grouped correctly for permanent archiving via the bookbinder. 

The detector readout system have their own servers where data is framed into .g3 files. The suprsync agent manages these files, ensuring they are transferred from the server to the site-managed DAQ node (in this case, \code{daq-satp1}) where the .g3 files are supposed to reside. For bookbinding, the stop time for the field \lstinline{daq-sync-smurf-so3.feeds.archive_stats.finalized_until} indicates the last time the \code{daq-sync-smurf-so3} suprsync agent successfully checked the files on the SMuRF server and copied them to \code{daq-satp1}, making them available for bookbinding. This finalized time and the streaming times from the pysmurf monitors are critical to the bookbinder.

\subsubsection{For Aggregate HK Properties}
%For Level 3 data, \texttt{\code{g3thk}} is currently being tested for site deployment to include info useful for querying instrument and agent states critical to understanding the telescope's status both during and outside observation mode. Many users rely on observation data, and so databases based on observation books exist and contain essential HK information. However, a database based on observation books leaves gaps in information when the telescope is not observing---information about HK data is still needed to ensure instrumentation and systematics are managed effectively. Instrument testing, PID testing, and calibration systematics require querying information from a larger dataset not covered by current observation databases, which often only extract a minimal set of HK data, such as PWV values, without fully contextualizing the full instrument.
%HK data loading

%\hk also contains data properties on all fields.  Can be used to investigate median (etc) quantities without loading any raw HK data. Can also be used to identify time periods for loading HK data

%summarizes HK instrument information per HK book
As noted above, \hk also functions as a data loader; it has also been developed for Level 3 data analysis, enabling efficient exploration of specific HK data sets. \hk uses Level 3 data to summarize instrument information (mean, median, minimum, maximum, and standard deviation) for each field in an HK file, allowing users to query specific properties without reading entire files. At Level 3, \hk focuses on a reduced number of agents. In a SAT-MF1 \hk instance, $\sim$14 of SAT-MF1's 69 agents are saved per HK file, reducing the number of fields from $\sim$1500 to $\sim$400. This subset includes only essential agents needed to diagnose the telescope's cryogenic, calibrator, and platform statuses, such as Lakeshore 240s, Lakeshore 372s, Bluefors dilution refrigerator, half-wave plate encoder, wiregrid encoder, and the ACU. By excluding hundreds of fields from power supply agents and core \code{ocs} agents like the registry agent, it prevents the database from growing excessively large at a rapid rate.  Access to this data is critical for hardware experts to debug and fine-tune telescope parameters, as well as to explore systematics, as the database continuously collects information regardless of whether an observation is occurring. For example, one might investigate the median focal plane temperature from a Lakeshore 372 in relation to scan speed from the ACU using the values provided in the database.

\begin{figure}[ht]
    \centering
    \includegraphics[width = \linewidth]{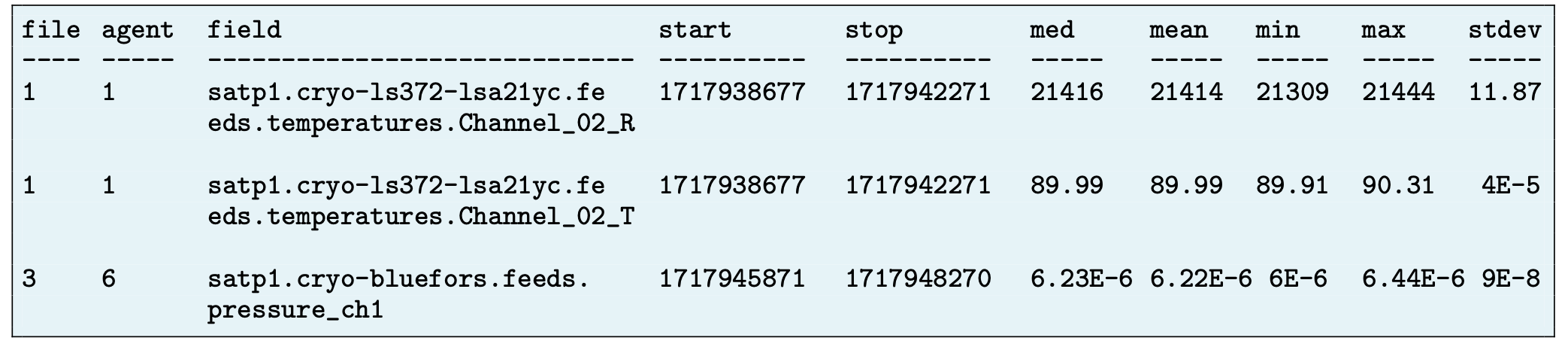}
    \caption{Level 3 SAT-MF1 \hk example of HK fields. Here, aggregate properties are provided for 2 fields from a Lakeshore 372 device (temperature and resistance for channel 2) and a Bluefors dilution refrigerator pressure sensor (for channel 1). These aggregate properties are calculated per HK file, and can be used for quick analysis without the need to explore .g3 files. However, they were most commonly used to load data under specific conditions such as when the scan speed of the telescope was a specific value and the maximum focal plane temperature was $\sim$95K, for example.}
    \label{fig:hksat}
\end{figure}

\begin{figure}[ht]
    \centering
    \includegraphics[width = \linewidth]{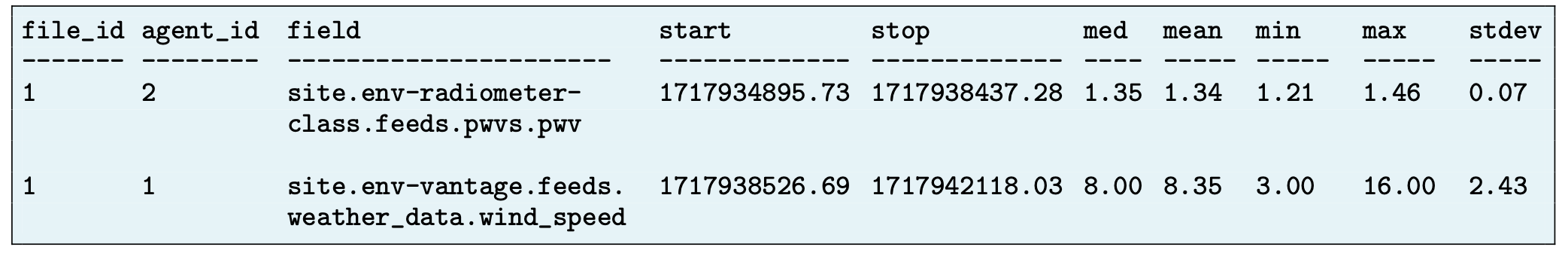}
    \caption{Level 3 site \hk example where the top entry describes atmospheric water vapor readings from the site radiometer, and the bottom field identifies wind speed from the weather monitor. These parameters, especially PWV, can be used in conjunction with other \hk instances to refine data exploration. }
    \label{fig:hksite}
\end{figure}

Note that multiple databases can be queried to improve and refine data exploration. For example, a common query involves checking the telescope's azimuth position using the SAT-MF1 \hk instance when atmospheric water vapor conditions were less than 2mm, which requires querying the site \hk instance. This type of query can cover a time range from a few weeks to several months, returning data only from files that meet the specified conditions. With SAT-MF1 being one of two mid-frequency SATs deployed for SO, \hk is also highly beneficial for quickly expanding analysis across telescopes under specific conditions, enabling efficient cross-analysis.

% To Do:

%azimuth and elevation positions of the telescope based on specific parameters, like periods when the precipitable water vapor (PWV) was less than 2mm.  

%provide list of agents (quantity of too long) that would be used for g3thk 
%

\subsubsection{HK Data Loading}

Loading SO HK data is cumbersome due to the serialized, asynchronous data structure, which provides flexibility but results in a non-seekable format. Data sources can be sporadic (e.g., calibration in the lab) or consistently present (e.g., cold and warm thermometry). 
\hk's data loader was developed for lab use, but its reading time increases with the number of .g3 files. Consequently, after five months of commissioning data from SAT-MF1, data loading became prohibitively slow. 

To address this, a new data loading mechanism called \texttt{hkdb} was developed to replace the one most commonly used on the backend of \hk, enabling fast and efficient exploration of HK data for the detector analysis pipeline. This mechanism uses a supplemental database to index each .g3 frame. For each data frame, this index tracks the .g3 file in which it is located, the byte offset, start and stop time of the frame, the \code{ocs} agent instance-id, and the feed\cite{ocs} name writing the data. At the site, this index is implemented using PostgreSQL for compatibility with the NFS and is deployed per telescope and the site system.

\lstset{
  basicstyle=\ttfamily\small,
  numberstyle=\tiny,
  stepnumber=1,
  numbersep=5pt,
  backgroundcolor=\color{cyan!10}, %lime!18
  frame=single,
  aboveskip=5pt,
  belowskip=5pt,
  lineskip=-1pt,
  showstringspaces=false,
  columns=flexible,
  breaklines=true,
  breakatwhitespace=true,
  tabsize=4,
  prebreak=\mbox{\textcolor{red}{$\hookrightarrow$}\space},
  postbreak=\mbox{\textcolor{red}{$\hookrightarrow$}\space}
}

\section{Conclusion}

This proceedings described the deployment of \code{ocs} for SAT-MF1, the first of SO's SATs commissioned and observing at the Atacama site. We provided an overview of the software, nodes, and host architecture at the site, as well as the data characteristics for the receiver. We also discussed a smaller subset of the site software system that enables successful telescope operations for SAT-MF1. This subset also includes context on the use of open-source software hosted on the site system, ensuring compatibility across all telescopes. For SAT-MF1, \code{ocs} successfully acquires data from 69 agents with a total of 1524 HK fields at a rate of approximately 3.58 Mbps, as well as detector data from 7 wafers totaling 12,000 detectors. 

%\code{ocs} also provides precise coordination of commanding hardware components over their network interfaces or via direct connections to physical computers on the platform.

We also described the data access software designed for efficiently exploring housekeeping data to contextualize detector behavior, eliminating the tedious task of arbitrarily loading .g3 files. We concluded with a description of HK data reading tools for both data packaging and providing aggregate information for a HK file, as well as the development of an improved HK data loading tool.

As of this writing, \code{hkdb} has been deployed to the site and is transitioning to become the default data loading mechanism. One potential improvement for \hk is to use \code{hkdb} to replace its existing data loader, allowing for quick database queries and data loading without excluding any agents for Level 3 data. This ensures that information about any agent can be accessed when needed, even if queried rarely. Another feature to consider is expanding the database to include observation-level summary information, making it easier to search the database using observation IDs while also retaining the ability to explore data outside of observation periods. Finally, there may be value in converting \hk into a PostgreSQL database for compatibility with the NFS.

\clearpage
\section{Appendix A. Enclosures Node Setup}

\renewcommand{\arraystretch}{1.1}
\begin{table}[H]
    \centering
    \begin{tabular}{|p{5cm}|l|p{8cm}|}
    \hline
    \textbf{Location} & \textbf{Node} & \textbf{Description} \\
    \hline
    Azimuth Enclosure (SAT-MF1 Platform) & NUC (Linux) & Contains network and power interfaces to the observatory site. \\
    \hline
    DR Gas Handling System (SAT-MF1 Platform) & None & Contains the gas-handling system for the cryostat’s DR insert, and includes electrical, vacuum, helium mixture, compressed air, and network interfaces. \\
    \hline
    Housekeeping Enclosure 1 (SAT-MF1 Platform) & None & Housekeeping 1 only contains two Lakeshore 372s for cryogenic thermometry readout and heater control for dilution refrigerator. \\
    \hline
    Housekeeping Enclosure 2 (SAT-MF1 Platform) & None & Housekeeping 2 contains 4 Lakeshore 240s, network interfaces and multifunction DAQ devices. \\
    \hline
    ATCA Crates (SAT-MF1 Platform) & None & The SMuRF electronics for detector readout, of which there are 2.   \\
    \hline
    Coolant Control (SAT-MF1 Platform) & IFM IO-Link & Provides coolant flow to the ATCA crates as well as telescope receiver components. IFM IO-Link used to connect to the flowmeters in the enclosure to read out coolant tempearature and flow.\\
    \hline
    CHWP Gripper Enclosures (SAT-MF1 Platform) & None & Together, they contain the 3 actuators for the cryostat’s cold half-wave plate grippers, and include only electrical interfaces. \\
    \hline
    CHWP and Grid Loader Enclosure (SAT-MF1 Platform) & NUC (Linux) & Contains the electronics to operate and read out both the spinning cold half wave plate, and the sparse wire grid loader, and includes electrical and network interfaces. \\
    \hline
    Turbo Pump Enclosure (SAT-MF1 Platform) & None & Contains 2 turbo pumps used by the DR GHS to operate the cryostat's DR; includes electrical, vacuum, helium mixture, compressed air, and network interfaces. \\ 
    \hline
    Site Office & daq-satp1 & Site-managed daq node for SAT-MF1 for Agent Management. Is the main node, where 36 of the 69 SAT-MF1 OCS Agents live. \\ 
    \hline
    Site Office & SMuRF Servers & Run the Agents and SMuRF-related software to readout SAT-MF1 detector arrays. \\
    \hline
    Site Container & NUC (Windows) & One of several containers at the site; this particular one houses the NUC that interfaces with SAT-MF1's dilution refrigerator. The Bluefors NUC is located on a rack near the large ACU control unit in this container which is used to communicate with SAT-MF1's telescope platform. \\
    \hline    
    \end{tabular}
    \vspace{0.5mm}
    \caption{Table listing the SAT-MF1 enclosure setup at the site, detailing the purpose of each enclosure and identifying those that contain nodes for network connections that cannot be supported directly by the \code{daq-satp1} node in the site office.}
    \label{tab:enclosures}
\end{table}

\newpage

\section*{ACKNOWLEDGMENTS} % equivalent to \section*{ACKNOWLEDGMENTS}       

This work was supported in part by a grant from the Simons Foundation (Award \#457687, B.K.). This work was also supported by the National Science Foundation (UEI GM1XX56LEP58). %We would like to thank the communities of the many open-source packages in use with \code{campana} and \code{sotodlib}

% References
\bibliography{main} % bibliography data in main.bib

\begin{thebibliography}{10}

\bibitem{sogoals}
{The Simons Observatory Collaboration}, ``The simons observatory: science goals and forecasts,'' {\em Journal of Cosmology and Astroparticle Physics}~{\bf 2019},  056--056 (feb 2019).

\bibitem{alarms}
Nguyen, D.~V., ``{The Simons Observatory: Alarms and Detector Quality Monitoring},'' in [{\em International Society for Optics and Photonics}{\nolinebreak\hspace{0.1em}]},  SPIE (in press).

\bibitem{ocs}
Koopman, B.~J., Lashner, J., Saunders, L.~J., Hasselfield, M., Bhandarkar, T., Bhimani, S., Choi, S.~K., Duell, C.~J., Galitzki, N., Harrington, K., Hincks, A.~D., Ho, S.-P.~P., Newburgh, L., Reichardt, C.~L., Seibert, J., Spisak, J., Westbrook, B., Xu, Z., and Zhu, N., ``{The Simons Observatory: overview of data acquisition, control, monitoring, and computer infrastructure},'' in [{\em Software and Cyberinfrastructure for Astronomy VI}{\nolinebreak\hspace{0.1em}]},  Guzman, J.~C. and Ibsen, J., eds.,  {\bf 11452},  1145208, International Society for Optics and Photonics, SPIE (2020).

\bibitem{galitzki2024simons}
Galitzki, N., Tsan, T., Spisak, J., Randall, M., Silva-Feaver, M., Seibert, J., Lashner, J., Adachi, S., Adkins, S.~M., Alford, T., Arnold, K., Ashton, P.~C., Austermann, J.~E., Baccigalupi, C., Bazarko, A., Beall, J.~A., Bhimani, S., Bixler, B., Coppi, G., Corbett, L., Crowley, K.~D., Crowley, K.~T., Day-Weiss, S., Dicker, S., Dow, P.~N., Duell, C.~J., Duff, S.~M., Gerras, R.~G., Groh, J.~C., Gudmundsson, J.~E., Harrington, K., Hasegawa, M., Healy, E., Henderson, S.~W., Hubmayr, J., Iuliano, J., Johnson, B.~R., Keating, B., Keller, B., Kiuchi, K., Kofman, A.~M., Koopman, B.~J., Kusaka, A., Lee, A.~T., Lew, R.~A., Lin, L.~T., Link, M.~J., Lucas, T.~J., Lungu, M., Mangu, A., McMahon, J.~J., Miller, A.~D., Moore, J.~E., Morshed, M., Nakata, H., Nati, F., Newburgh, L.~B., Nguyen, D.~V., Niemack, M.~D., Page, L.~A., Sakaguri, K., Sakurai, Y., Rao, M.~S., Saunders, L.~J., Shroyer, J.~E., Sugiyama, J., Tajima, O., Takeuchi, A., Bua, R.~T., Teply, G., Terasaki, T., Ullom, J.~N., Lanen, J. L.~V., Vavagiakis, E.~M.,
  Vissers, M.~R., Walters, L., Wang, Y., Xu, Z., Yamada, K., and Zheng, K., ``The simons observatory: Design, integration, and testing of the small aperture telescopes,'' (2024).

\bibitem{Zhu_2021}
Zhu, N., Bhandarkar, T., Coppi, G., Kofman, A.~M., Orlowski-Scherer, J.~L., Xu, Z., Adachi, S., Ade, P., Aiola, S., Austermann, J., Bazarko, A.~O., Beall, J.~A., Bhimani, S., Bond, J.~R., Chesmore, G.~E., Choi, S.~K., Connors, J., Cothard, N.~F., Devlin, M., Dicker, S., Dober, B., Duell, C.~J., Duff, S.~M., Dünner, R., Fabbian, G., Galitzki, N., Gallardo, P.~A., Golec, J.~E., Haridas, S.~K., Harrington, K., Healy, E., Ho, S.-P.~P., Huber, Z.~B., Hubmayr, J., Iuliano, J., Johnson, B.~R., Keating, B., Kiuchi, K., Koopman, B.~J., Lashner, J., Lee, A.~T., Li, Y., Limon, M., Link, M., Lucas, T.~J., McCarrick, H., Moore, J., Nati, F., Newburgh, L.~B., Niemack, M.~D., Pierpaoli, E., Randall, M.~J., Sarmiento, K.~P., Saunders, L.~J., Seibert, J., Sierra, C., Sonka, R., Spisak, J., Sutariya, S., Tajima, O., Teply, G.~P., Thornton, R.~J., Tsan, T., Tucker, C., Ullom, J., Vavagiakis, E.~M., Vissers, M.~R., Walker, S., Westbrook, B., Wollack, E.~J., and Zannoni, M., ``The simons observatory large aperture telescope
  receiver,'' {\em The Astrophysical Journal Supplement Series}~{\bf 256},  23 (Sept. 2021).

\bibitem{Gudmundsson_2021}
Gudmundsson, J.~E., Gallardo, P.~A., Puddu, R., Dicker, S.~R., Adler, A.~E., Ali, A.~M., Bazarko, A., Chesmore, G.~E., Coppi, G., Cothard, N.~F., Dachlythra, N., Devlin, M., Dünner, R., Fabbian, G., Galitzki, N., Golec, J.~E., Patty~Ho, S.-P., Hargrave, P.~C., Kofman, A.~M., Lee, A.~T., Limon, M., Matsuda, F.~T., Mauskopf, P.~D., Moodley, K., Nati, F., Niemack, M.~D., Orlowski-Scherer, J., Page, L.~A., Partridge, B., Puglisi, G., Reichardt, C.~L., Sierra, C.~E., Simon, S.~M., Teply, G.~P., Tucker, C., Wollack, E.~J., Xu, Z., and Zhu, N., ``The simons observatory: modeling optical systematics in the large aperture telescope,'' {\em Applied Optics}~{\bf 60},  823 (Jan. 2021).

\bibitem{McCarrick_2021}
McCarrick, H., Healy, E., Ahmed, Z., Arnold, K., Atkins, Z., Austermann, J.~E., Bhandarkar, T., Beall, J.~A., Bruno, S.~M., Choi, S.~K., Connors, J., Cothard, N.~F., Crowley, K.~D., Dicker, S., Dober, B., Duell, C.~J., Duff, S.~M., Dutcher, D., Frisch, J.~C., Galitzki, N., Gralla, M.~B., Gudmundsson, J.~E., Henderson, S.~W., Hilton, G.~C., Ho, S.-P.~P., Huber, Z.~B., Hubmayr, J., Iuliano, J., Johnson, B.~R., Kofman, A.~M., Kusaka, A., Lashner, J., Lee, A.~T., Li, Y., Link, M.~J., Lucas, T.~J., Lungu, M., Mates, J. A.~B., McMahon, J.~J., Niemack, M.~D., Orlowski-Scherer, J., Seibert, J., Silva-Feaver, M., Simon, S.~M., Staggs, S., Suzuki, A., Terasaki, T., Thornton, R., Ullom, J.~N., Vavagiakis, E.~M., Vale, L.~R., Van~Lanen, J., Vissers, M.~R., Wang, Y., Wollack, E.~J., Xu, Z., Young, E., Yu, C., Zheng, K., and Zhu, N., ``The simons observatory microwave squid multiplexing detector module design,'' {\em The Astrophysical Journal}~{\bf 922},  38 (Nov. 2021).

\bibitem{ocs24}
Koopman, B.~J., ``{The Simons Observatory: Deployment of the Observatory Control System and supporting infrastructure},'' in [{\em Software and Cyberinfrastructure for Astronomy VIII}{\nolinebreak\hspace{0.1em}]},  International Society for Optics and Photonics, SPIE (in press).

\bibitem{smurf}
Henderson, S.~W., Ahmed, Z., Austermann, J., Becker, D., Bennett, D.~A., Brown, D., Chaudhuri, S., Cho, H.-M.~S., D'Ewart, J.~M., Dober, B., Duff, S.~M., Dusatko, J.~E., Fatigoni, S., Frisch, J.~C., Gard, J.~D., Halpern, M., Hilton, G.~C., Hubmayr, J., Irwin, K.~D., Karpel, E.~D., Kernasovskiy, S.~S., Kuenstner, S.~E., Kuo, C.-L., Li, D., Mates, J. A.~B., Reintsema, C.~D., Smith, S.~R., Ullom, J., Vale, L.~R., Winkle, D. D.~V., Vissers, M., and Yu, C., ``{Highly-multiplexed microwave SQUID readout using the SLAC Microresonator Radio Frequency (SMuRF) electronics for future CMB and sub-millimeter surveys},'' in [{\em Millimeter, Submillimeter, and Far-Infrared Detectors and Instrumentation for Astronomy IX}{\nolinebreak\hspace{0.1em}]},  Zmuidzinas, J. and Gao, J.-R., eds.,  {\bf 10708},  1070819, International Society for Optics and Photonics, SPIE (2018).

\bibitem{Kernasovskiy_2018}
Kernasovskiy, S.~A., Kuenstner, S.~E., Karpel, E., Ahmed, Z., Van~Winkle, D.~D., Smith, S., Dusatko, J., Frisch, J.~C., Chaudhuri, S., Cho, H.~M., Dober, B.~J., Henderson, S.~W., Hilton, G.~C., Hubmayr, J., Irwin, K.~D., Kuo, C.~L., Li, D., Mates, J. A.~B., Nasr, M., Tantawi, S., Ullom, J., Vale, L., and Young, B., ``Slac microresonator radio frequency (smurf) electronics for read out of frequency-division-multiplexed cryogenic sensors,'' {\em Journal of Low Temperature Physics}~{\bf 193},  570–577 (May 2018).

\bibitem{nextline}
Sakuma, T., Guan, Y., Hasselfield, M., Koopman, B., Newburgh, L., and Nguyen, D., ``Nextline.'' https://doi.org/10.5281/zenodo.11451619 (2024).

\bibitem{yilun_spie}
Guan, Y., ``{Simons Observatory: Observatory Scheduler and Automated Data Processing},'' in [{\em Software and Cyberinfrastructure for Astronomy VIII}{\nolinebreak\hspace{0.1em}]},  International Society for Optics and Photonics, SPIE (in press).

\end{thebibliography}
\bibliographystyle{spiebib} % makes bibtex use spiebib.bst

\end{document}